\documentclass[onecolumn]{elsarticle}
\journal{}

\usepackage{rotating}
\usepackage{graphicx}
\usepackage{color}
\usepackage{amsmath}

\usepackage{tabularx}

\usepackage{caption}
\usepackage{subcaption}

\begin{document}
\begin{frontmatter}
\title{Material phase classification by means of Support Vector Machines}


\author[lab1]{Jaime~Ortegon}
\author[lab2]{Rene~Ledesma-Alonso}
\author[lab1]{Romeli~Barbosa}
\author[lab1]{Javier~V\'azquez~Castillo}
\author[lab3]{Alejandro~Castillo~Atoche}

\address[lab1]{Universidad de Quintana Roo, Boulevar Bah\'ia s/n, Chetumal, 77019, Quintana Roo, M\'exico}
\address[lab2]{CONACYT-Universidad de Quintana Roo, Boulevar Bah\'ia s/n, Chetumal, 77019, Quintana Roo, M\'exico}
\address[lab3]{Universidad Aut\'onoma de Yucat\'an, Av. Industrias no contaminantes s/n, M\'erida, 150, Yucat\'an, M\'exico}
\markboth{Journal of Computational Materials Science}%
{Ortegon et al.}

\begin{abstract}
The pixel's classification of images obtained from random heterogeneous materials is a relevant step to compute their physical properties, like Effective Transport Coefficients (ETC), during a characterization process as stochastic reconstruction.
A bad classification will impact on the computed properties; however, the literature on the topic discusses mainly the correlation functions or the properties formulae, giving little or no attention to the classification; authors mention either the use of a threshold or, in few cases, the use of Otsu's method.
This paper presents a classification approach based on Support Vector Machines (SVM) and a comparison with the Otsu's-based approach, based on accuracy and precision. The data used for the SVM training are the key for a better classification; these data are the grayscale value, the magnitude and direction of pixels’ gradient. 
For instance, in the case study, the accuracy of the pixel's classification is 77.6\% for the SVM method and 40.9\% for Otsu's method.
Finally, a discussion about the impact on the correlation functions is presented in order to show the benefits of the proposal.

\end{abstract}


\end{frontmatter}

\section{Introduction}

The study of heterogeneous materials is common in the literature, and it is useful to determine its effective transport coefficients, electromagnetic and mechanical properties.
Digital images are used to perform a characterization of materials based on morphological properties.
Coker and Torquato \cite{torquato1995} established a methodology to compute correlation functions with the purpose of describing the morphology of a digitazed binary material.
Recent works are devoted to reconstruct three-dimensional materials based on the correlation functions using different methods \cite{Pant2014,Pant2015,barbosa_stochastic_2011,Joos2011,Xu2014,Ender2011}.
However, most works use a threshold to convert gray-scale to binary images; the threshold may be fixed \cite{Xu2017}, automatically selected with Otsu's method \cite{Pant2014,Pant2015, Joos2011}, Maximum Likelihood \cite{Ender2011} or Maximum Entropy \cite{Zhu2014}.
Recently, Sabharwal et al. \cite{Sabharwal2016} use Sauvola's method to enhance the binarization \cite{Sauvola00}; however, they barely comment the difference with the results obtained with the Otsu's method.

For instance, Xu et al. \cite{Xu2014} denoised greyscale SEM images and applied a threshold.
The greyscale threshold is determined to match the actual volume fraction; they propose the material characterization using a small set of microstructure descriptors  for composition, dispersion status, and phase geometry, finishing with a 3D reconstruction based on 3D descriptors.
Xu, Gao and Li \cite{Xu2017}  only mention that is possible to denoise and binarize an image, and then to derive a polynomial expression, by means of a PCA (Principal Component Analysis)-based descriptor, used to reconstruct the material.
Pant, Mitra and Secanell \cite{Pant2014,Pant2015} use the Otsu's method to binarize an image, with the subsequent use of diverse heuristic methods, to improve the speed of the reconstructed material.
Likewise, Joos et al. \cite{Joos2011} used Otsu's method and compared its performance with that of a mean-algorithm and ``intuiteve '' inspection, exhibiting the importance of image processing and binarization; they also studied the differences between several parameters, such as the volume fraction and the tortuosity. 
Ender et al. \cite{Ender2011} propose the application of two thresholds, to get a trimodal segmentation, i.e., a maximum likelihood method is implemented to fit three normal distributions to the grayscale values of the image, from which the two equiprobable abcissae correspond to the two thresholds.
The Maximum Entropy in the phases is used by Zhu et al. \cite{Zhu2014} to determine a single threshold to binarize an image; afterwards, they use a mesh method to model an aluminum foam and to calculate its compression performance.
Sabharwal et al. \cite{Sabharwal2016} use Souvola's method to binarize an image and then they generate a 3D connected pore network (mesh) from a stack of 2-D images.

There are few examples of the employment of Support Vector Machines (SVMs) in materials science.
Sundararaghavan and Zabaras \cite{Sundararaghavan2005,Sundararaghavan2004} use SVMs to clasiffy 3D microstructures obtained computationally using a Monte-Carlo Potts grain growth model.
Recently, Ortegon et al. \cite{OrtegonAguilar2016} present the idea of using SVMs to binarize images; however, neither the presented procedure was validated nor the effects on the correlation functions were studied.

This article is organized as follows.
Section \ref{images} presents the image processing algorithms used.
Section \ref{svm} gives a brief on Support Vector Machines, while the classification procedure is presented on Section \ref{classification}.
In order to observe the effects on the correlation functions, they are described in section \ref{correlation} and Section \ref{resultss} is devoted to the corresponding results and discussion.
Finally, conclusions are given in Section \ref{conclusion}.

\section{Image processing}
\label{images}

Images are a main source of information for humans; they may represent scalar fields such as light intensity, temperature, reflectance and density, among other relevant variables for science.
In the present work Scanning Electron Microscope (SEM) images are considered as the source of information about the structure of random heterogeneous materials (RHMs).
Particularly, images of the Catalytic Layer (CL) of Proton Exchange Membrane Fuel Cells (PEMFC) are studied.
The pixels in each of those images are classified to generate a representation of the configuration of the material phases, from which some microstructure statistial descriptors, i.e., low order correlation functions, are computed.

\begin{figure}[h!]
\centering
\includegraphics[width=3in]{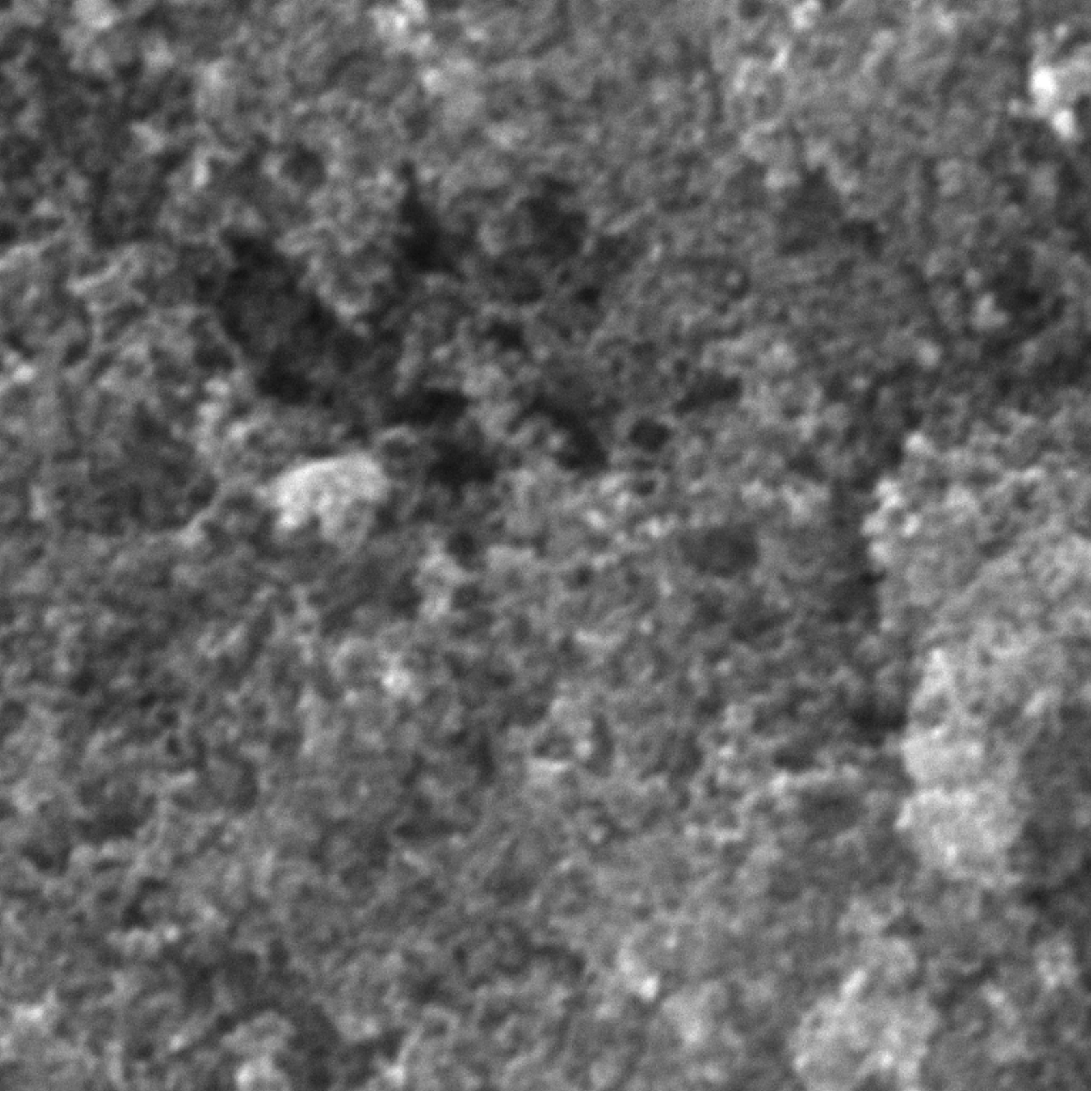}
\caption{SEM image of the catalytic layer of a PEMFC, which presents a bias on the first order statistic.}
\label{fig_dark}
\end{figure}

In computer science, images are represented as two or three dimensional arrays.
Usually, the first two dimensions represent the spatial position of a pixel, while the third dimension represents the channel (band) of the gathered information.
For example, a color image has three different bands (red, green and blue), whereas a gray scale image only has one band.
Image processing programs may use different data types to handle information, being the most widely used the ``unsigned char'', i.e., unsigned integers between 0 and 255.
Depending on the scene captured on the image, it is possible that the first order statistic of pixel values (minimum value of the grayscale observed on the image) is biased, as it is illustrated with the sample image shown in Figure \ref{fig_dark}.

\begin{figure}[h!]
\centering
	\includegraphics[width=3in]{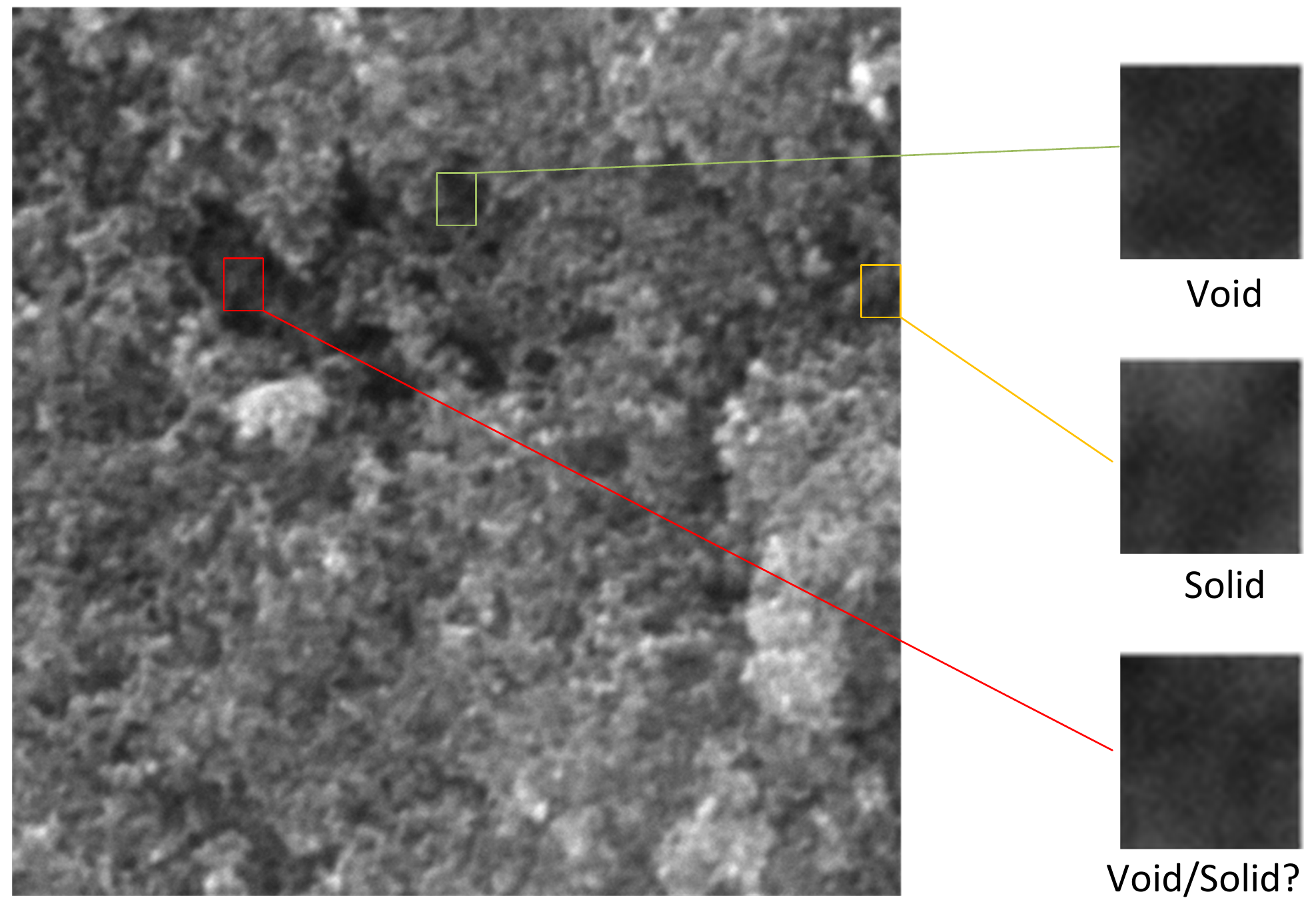}
	\caption{Close-up to regions with the same gray value corresponding to a different phase of the material.}
	\label{celdazoom}
\end{figure}

The images of two-phases RHMs are commonly separated in two classes.
This represents a binarization of the pixels, where black and white pixels represent void and solid phases of a porous material.
It is important to remember that the images are two-dimensional projections of a three-dimensional material, hence, it is possible that the same gray value corresponds to distinct parts of the material at different depths from the microscope sensor.
This situation has been presented in Figure \ref{celdazoom}, where pixels at different locations, either in the void phase or in the solid phase, display the same gray value.

\begin{figure}[h!]
\centering\begin{tabular}{ccc}
		\includegraphics[width=1.5in]{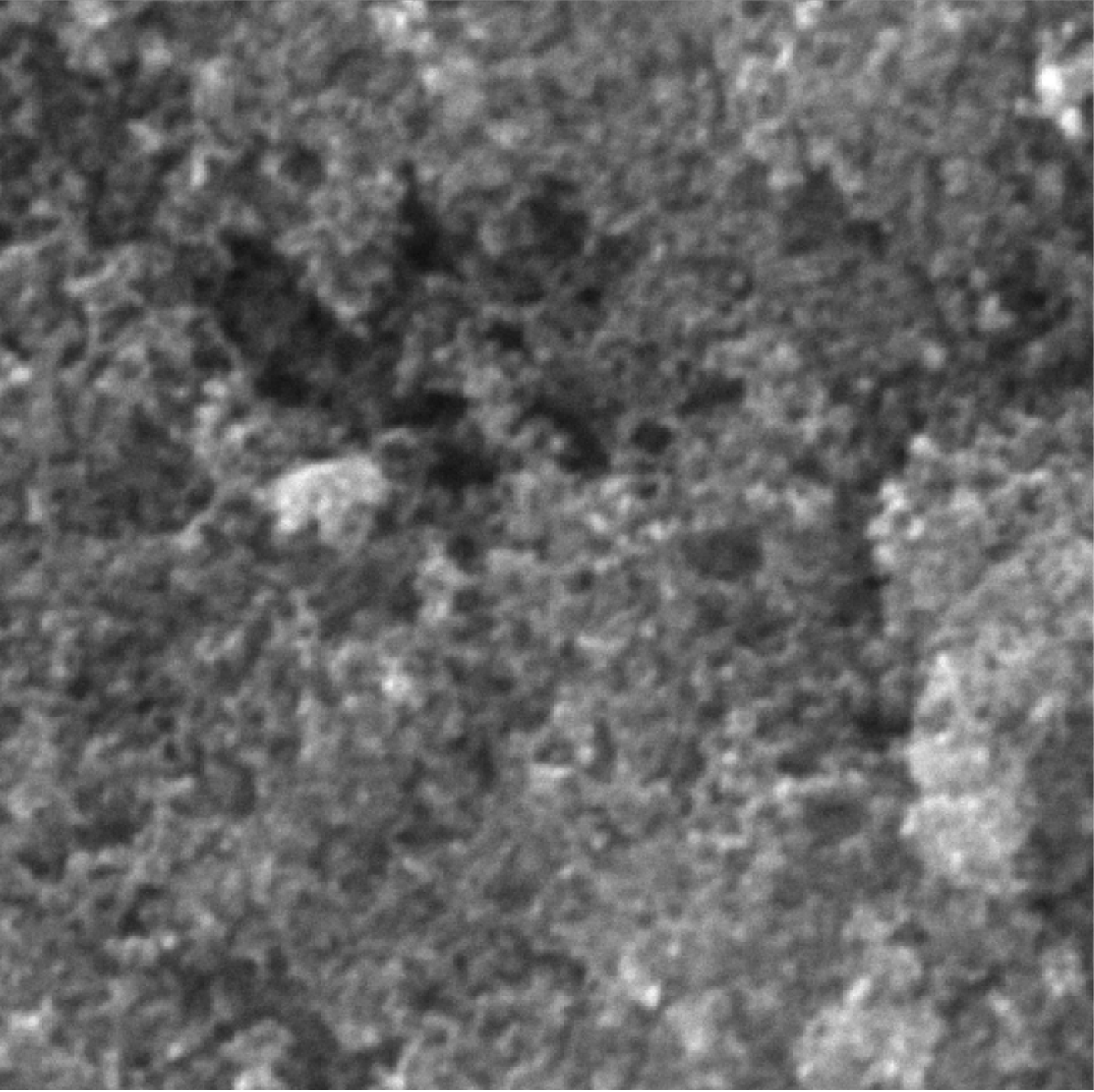} & &		\includegraphics[width=1.5in]{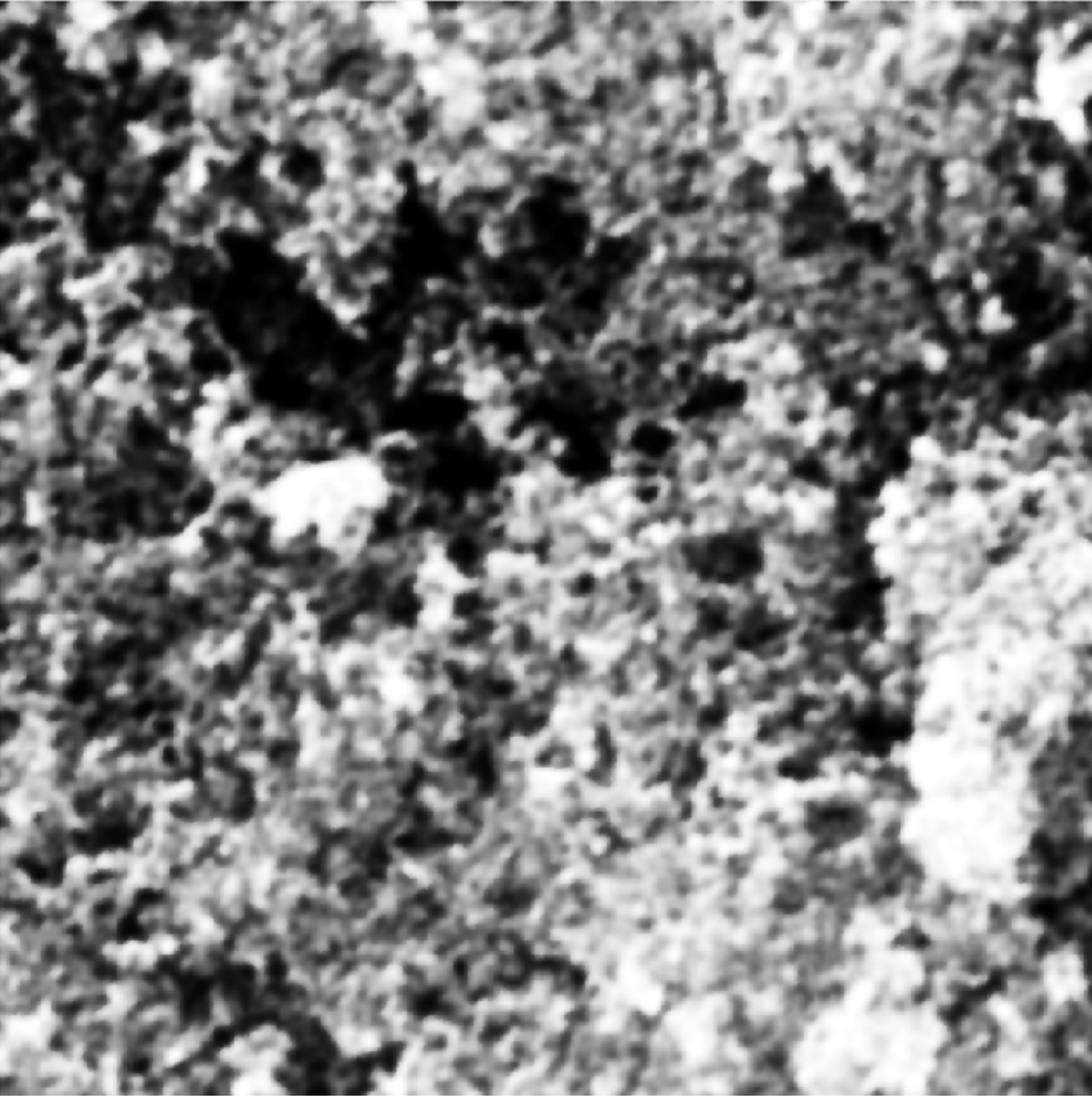} \\
	a) Original & & b) Enhanced
	\end{tabular}
	\caption{Catalytic layer of a PEMFC obtained by SEM. a) Original image and b) enhanced image by the equalization-normalization-denoising method.} 
	\label{original}
\end{figure}

SEM images are in some cases blurry or too dark to be easily analyzed by human eye.
In consequence, it is common to enhance the image with different methods like histogram equalization, normalization and denoising \cite{Gonzalez2008}.
Equalization is the process to distribute channel values of an image to the whole pixel's domain range, by spreading values and increasing the image contrast.
Images with high contrast shows a great distribution of gray-level detail.
Since different image formats may use different ranges for pixel's values, a normalization is proposed, in order to convert integers to floating point numbers between 0 and 1, which simplifies the classification.
Finally, denoising will remove gaussian and impulse noise with a Weiner and median filter respectively. 
Original and processed images, obtained with the aforementioned equalization-normalization-denoising method, are shown in Figure \ref{original}.

\section{Support Vector Machines}
\label{svm}

Support Vector Machines (SVMs) are based on hyperplane classifiers \cite{scholkopf_learning_2002}, making them suitable for problems which are linearly separable in some feature space.
SVMs are supervised machine learning algorithms; in this sense, they are kernel algorithms which use statistical learning theory.
Supervised learning have, at least, two kind of data: training and test.
Training data consist of a set of pairs, each one consisting of an input and a desired output, which are used by the SVM to learn an inferring function.
Test data are processed by the SVM to infer its output, which is compared to the desired output, in order to measure the accuracy of the SVM. 
The simplest case is to classify data considering only one class, with each datum belonging or not to such class.
SVMs aim to find ``the best'' hyperplane, among all hyperplanes separating the data, i.e., the one with the maximum margin of separation with respect to any training data.  

However, SVMs are useful even in the case of redinput data that are not linearly separable in the input space, because SVMs may first map the input space to a higher dimensional space, using a non linear function.
In such space, the input data are more likely to be linearly separable; nevertheless, it is possible that some inputs are not correctly separated, but this is acceptable in order to learn and perform better on new inputs.
Some common mappings are polynomial, sigmoid and other spline functions.

\begin{figure}[h!]
\centering
    \includegraphics[width=3in]{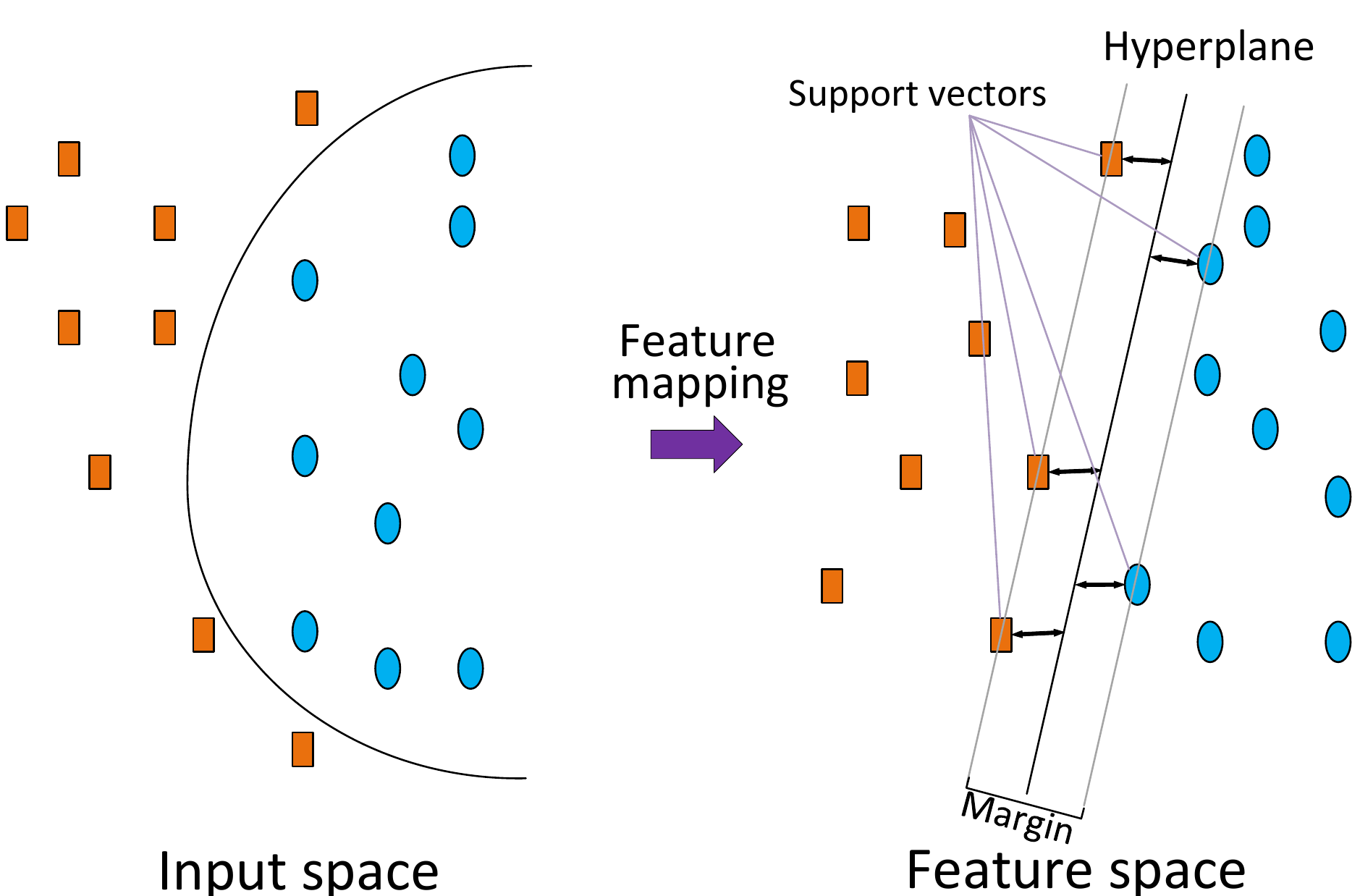}
	\caption{Schematic representation of a hyperplane separating two classes, found by a SVM.}
	\label{hiperplano}
\end{figure}

Summarizing, a SVM processes input data (possibly not linearly separable), maps them to a higher dimension space and finds a hyperplane to separate the data in two classes with the larger possible margin, as it is schematically represented in Figure \ref{hiperplano}.

\section{Phase Classification}
\label{classification}

The main feature of the approach presented in this manuscript is the phase classification performed by a SVM.
While other authors \cite{barbosa_stochastic_2011,das_effective_2009,jianjun_using_2011} use a simple threshold on the image's gray values, the proposed SVM methodology takes into account the image gradient (direction and magnitude), in addition to the grayscale field values.
The SVM allows the use of the three mentioned data as inputs, and it takes advantage of the additional data to improve the pixels classification.
Hence, the gradient, direction and magnitude, is computed for the pixels in the whole image.

During the training, one must provide the desired output to the SVM.
This is done manually by selecting and labeling sample regions of the image, which correspond to the void and solid phases.
The training data has 4 columns: gray value, gradient direction, gradient magnitude and phase label; each row corresponding to one pixel in the selected regions (void and solid phases).

\begin{figure}[h!]
\centering
	\begin{tabular}{ccc}
	\includegraphics[width=1.5in]{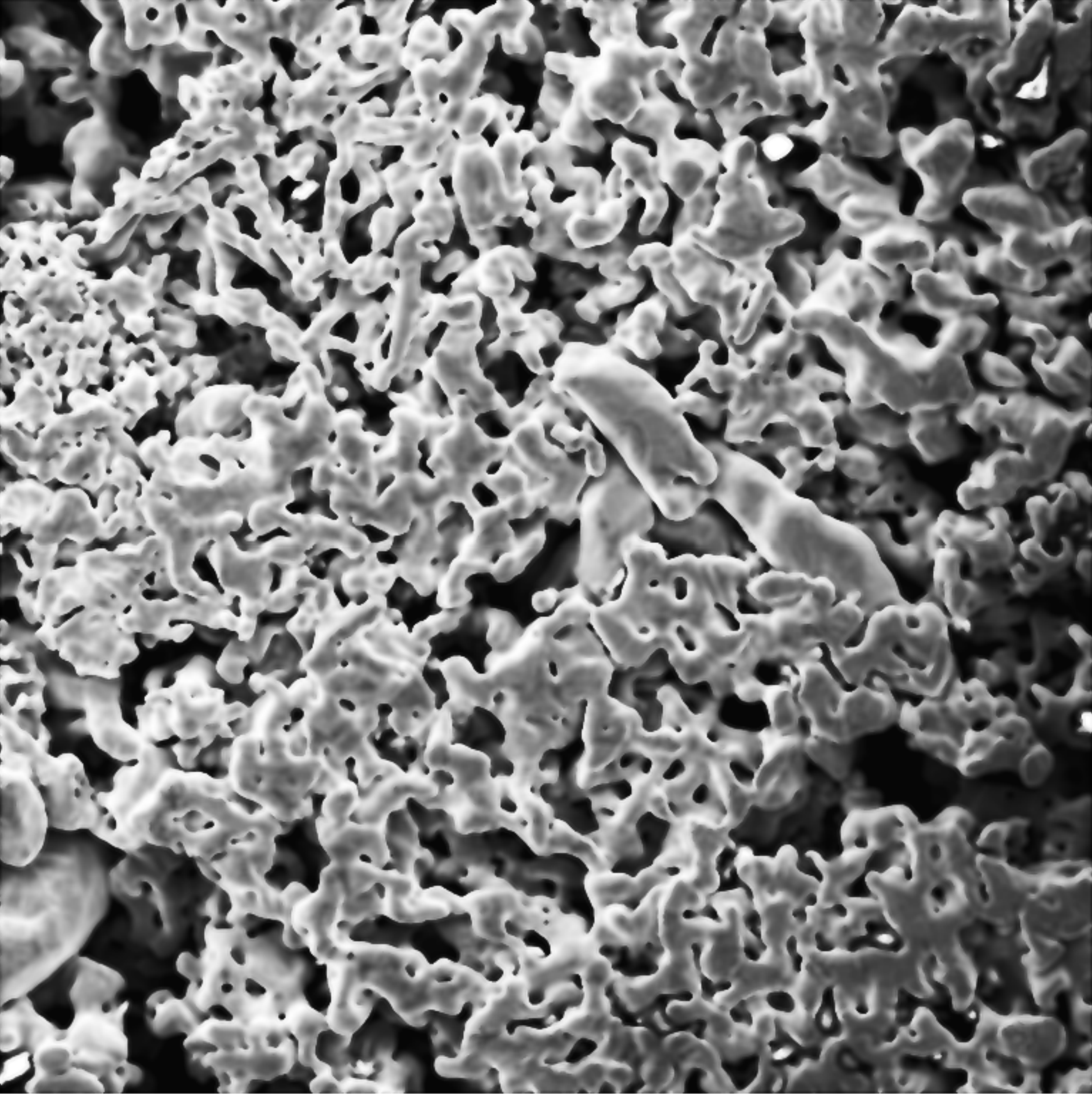} & &	\includegraphics[width=1.5in]{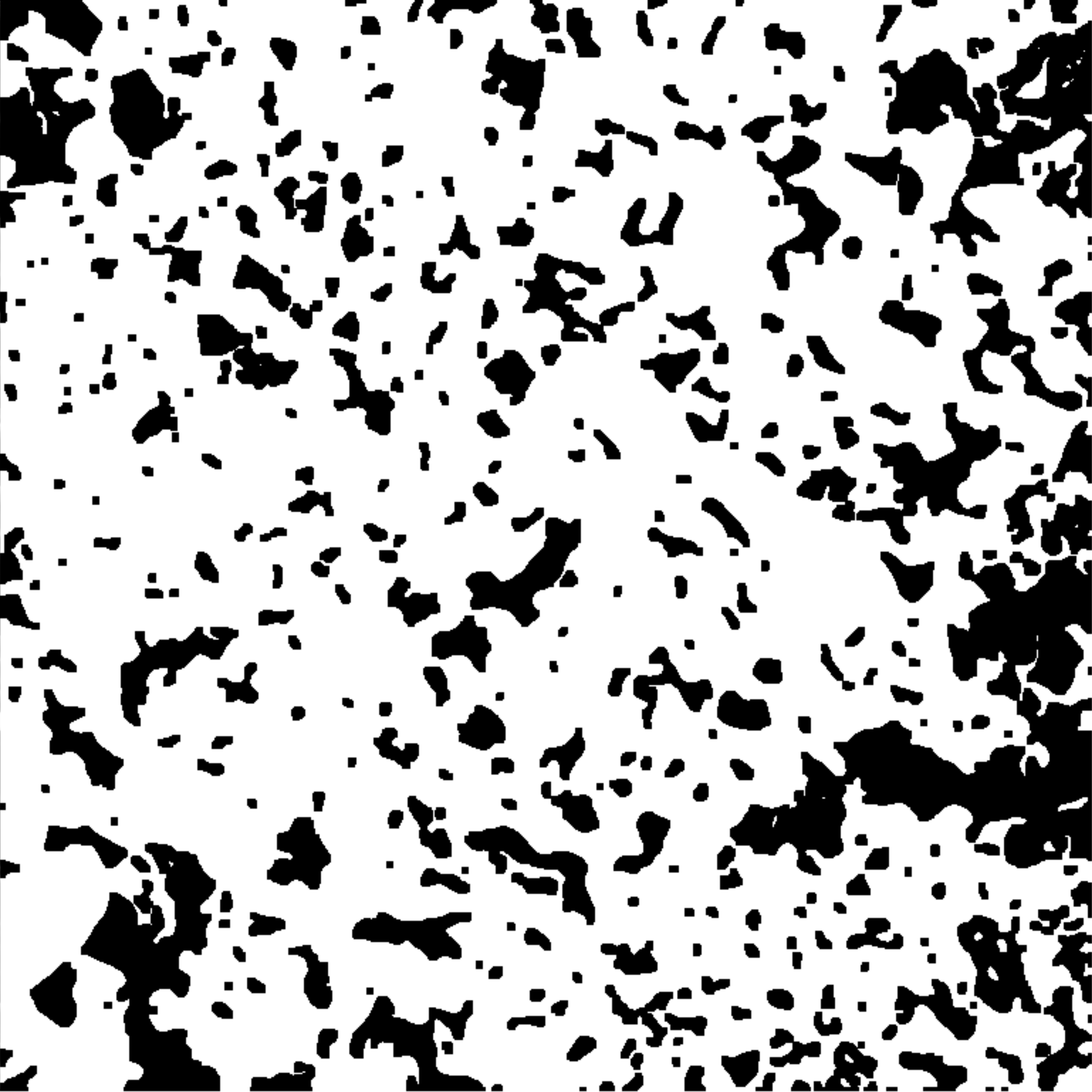} \\
	a) Original & & b) Binarization
	\end{tabular}
	\caption{Catalytic layer of a PEMFC obtained by SEM. a) Original image and b) binarization generated by means of the proposed SVM classification approach.}
	\label{binary}
\end{figure}

Once the SVM has been trainned, the whole image is processed, arranging the input data as a 3-column matrix with as many rows as pixels in the image.
The result of the SVM classification is a vector with the corresponding labels for each pixel.
After reshaping this vector to the image size, a binary image is obtained, where the solid phase is colored in white and void phase in black.
Figure \ref{binary} shows an example of an original image, a CL of a PEMFC, and the resulting binary image from the phase classification by means of SVM. 

\section{Correlation functions}
\label{correlation}

With the purpose of characterizing statistically the micro-structure of an heterogeneous material, as our first step, we define indicator functions for each phase $j\in\left[0,J-1\right]$, where $J$ is the total number of phases and $j=0$ indicates the void phase. For instance, for the phase $j$, the indicator function at the point described by $\mathbf{x}$ is defines as:
\begin{equation}
I_j(\mathbf{x})=\begin{cases}
1 & \text{ if } \mathbf{x} \text{ lies on phase $j$ } \\
0 & \text{ otherwise }
\end{cases} \ .
\end{equation}

The two-point correlation functions $S^{(2)}_j(\mathbf{x}_a,\mathbf{x}_b)$ is the probability that the two end points $\mathbf{x}_a$ and $\mathbf{x}_b$ of a line segment, with length $r=\vert \mathbf{x}_b-\mathbf{x}_a\vert$, fall in the same phase $j$:
\begin{equation}
S^{(2)}_j(\mathbf{x}_a,\mathbf{x}_b)\equiv
\Big\langle I_j\left(\mathbf{x}_a\right)I_j\left(\mathbf{x}_b\right)\Big\rangle \ ,
\end{equation}
where $\langle\cdot\rangle$ represents the expected value of its argument. 
For statistically homogeneous and isotropic media, the two-point correlation function does not depend on the specific positions of the end points, but only on the distance between them:
\begin{equation}
S^{(2)}_j(\mathbf{x}_a,\mathbf{x}_b)\rightarrow S^{(2)}_j(r)\equiv
\Big\langle I_j\left(\mathbf{x}\right)I_j\left(\mathbf{x}+\mathbf{r}\right)\Big\rangle \ ,
\end{equation}
where $\mathbf{r}$ is a vector of relative position with magitude $r$.

The line-path correlation function $L_j(\mathbf{x}_a,\mathbf{x}_b)$ is the probability that a line segment with end points $\mathbf{x}_a$ and $\mathbf{x}_b$, with length $r=\vert \mathbf{x}_b-\mathbf{x}_a\vert$, lies entirely on phase $j$:
\begin{equation}
L_j(\mathbf{x}_a,\mathbf{x}_b)\equiv 
\left\langle\left\lfloor\int_{0}^{1}I_j(\mathbf{x}_a+\alpha\left[\mathbf{x}_b-\mathbf{x}_a\right])d\alpha
\right\rfloor\right\rangle
 \ ,
\end{equation}
where $\alpha$ is an integration variable.
Again, for statistically homogeneous and isotropic media, the line-path correlation function does not depends anymore on the specific coordinates of $\mathbf{x}_a$ and $\mathbf{x}_b$.
Under this circumstances, this correlation function is only a function of the length $r$ of the line segment:
\begin{equation}
L_j(\mathbf{x}_a,\mathbf{x}_b)\rightarrow L_j(r)\equiv 
\left\langle\left\lfloor\int_{0}^{1}I_j\left(\mathbf{x}+\alpha\mathbf{r}\right)d\alpha\right\rfloor\right\rangle \ ,
\end{equation}
where $\mathbf{r}$ is a vector of relative position with magnitude r and an arbitrary direction.

\section{Results and discussion}
\label{resultss}
The proposed SVM classification is compared with Otsu's method, since most studies available in the literature use it to select the binarization threshold \cite{Pant2014,Pant2015,barbosa_stochastic_2011,Joos2011,Xu2014,Ender2011}. 
The image depicted in Figure \ref{binary} has been processed with both methods, selecting a lower right region to compare the results. 
Consequently, Figure \ref{binary_reg} presents a clipped region from a SEM image of $125\times 120$ pixels; on the left-hand side, the Otsu's binarization is shown, at the center, the original image is displayed and on the right-hand side, one can observe the SVM binarization. 

\begin{figure}[h!]
	\centering
	\begin{tabular}{ccccc}
		\includegraphics[width=1in]{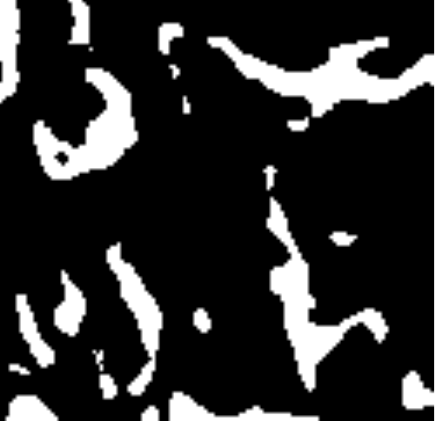} & &
		\includegraphics[width=1in]{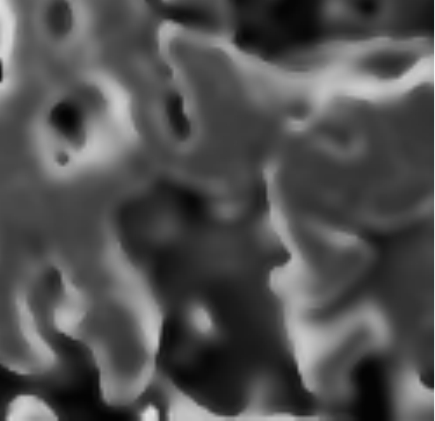} & &
		\includegraphics[width=1in]{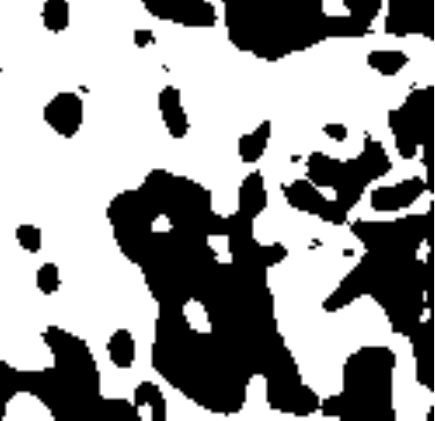} \\
		a) Otsu	& &
		b) Original & &
		c) SVM
	\end{tabular}
	\caption{a),c) Binarizations and b) original image, of a clipped region of $125\times 120$ pixels, obtained from the lower right corner of the image presented in Figure \ref{binary}.}
	\label{binary_reg}
\end{figure} 

In order to compare the results with those obtained with traditional methods, it is necessary to define concepts related with the expected and the predicted (obtained) results.
The phase classification is a 1-class problem, i.e., a pixel belongs or not to the solid class.
When a pixel is within the class, it takes a positive state, otherwise it takes a negative state; moreover, the expected result is named a \emph{real positive/negative} and the result obtained with the classification method is named a \emph{predicted positive/negative}.
In addition, when the state of the real (positive/negative) coincides with the state of the predicted (positive/negative), the combination of the two values gives a true state (positive/negative), otherwise the combination gives a false state (positive/negative).
For a better understanding of the above-mentioned states, the four combinations of real and predicted values are depicted in Table \ref{real-predicted}.

\begin{table}[h!]
	\centering
	\caption{Combinations of real and predicted results.}
	\label{real-predicted}
	\begin{tabular}{c|cc}
		&Real positives &Real negatives\\
		\hline
		Predicted positives&True positives (TP) &False positives (FP) \\
		Predicted negatives&False negatives (FN) &True negatives (TN)
	\end{tabular}
\end{table}

The real values were determined by a human being, who manually classified each pixel in the selected region.
Table \ref{true-false} presents the $TP$, $TN$, $FP$ and $FN$ values, according to the nomenclature given in Table \ref{real-predicted}, for the Otsu and SVM approach, and the number of total samples $TS=15,000$, which corresponds to the total number of pixels in the original image (displayed in Figure \ref{binary_reg}). 

\begin{table}[h!]
	\centering
	\caption{$TP$, $TN$, $FP$, $FN$ values for Otsu and SVM, $TS=15,000$. }
	\label{true-false}
	\begin{tabular}{l | rrrr}
		Method&$TP$ &$FP$ &	$TN$ & $FN$\\
		\hline
		Otsu&2509&66&3623&8802 \\
		SVM&8429&471&3218&2882
	\end{tabular}
\end{table}

The metrics used for the results are the accuracy, precision and recall, defined as follows:
\begin{subequations}
\begin{eqnarray}
Accuracy & = & (TP + TN) / (TS) \ , \\
Precision & = & TP/ (TP+ FP) \ , \\
Recall & =  &TP/ (TP+ FN) \ .
\end{eqnarray}
\label{eqs_metrics}
\end{subequations}
Table \ref{results} shows the metrics for each method, where the accuracy represents the fraction of the correctly predicted values.
In the abovementioned table, one observes that the SVM shows a better perfonce than Otsu's, with a value of 77.6\%.
The precision is the fraction of the correctly predicted positive values, for which the Otsu's approach shows a slightly better performance.
In this metric, Otsu's preponderance is misleading, since it is a consequence of the low value of the total positive values (2575 samples) predicted by this method, whereas for the SVM approach, the account rises to 8900 samples.
Finally, the recall measures the capacity of the method to reassemble the real positive values.
In this metric, the SVM can recall up to 74.5\% of the real positive values versus the 22.1\% of the Otsu's method.

\begin{table}[h!]
	\centering
	\caption{Results from the metrics, described by eqs.~\eqref{eqs_metrics}, applied to the binary images generated by the Otsu and SVM methods.}
	\label{results}
	\begin{tabular}{l | lll}
		Method&Accuracy&Precision&Recall\\
		\hline
		Otsu&0.4088&0.9744&0.2218 \\
		SVM&0.7765&0.9471&0.7452 
	\end{tabular}
\end{table}

\begin{table}[h!]
\centering
\caption{Image labels, properties and descriptions of the SEM images obtained from Catalytic Layers of PEMFCs. All the images present a pixel resolution of 768$\times$768.}
\label{Tab:Materials}
\begin{tabular}{c|c|p{9.5cm}}
Label & Magnification & Sample description \\
\hline
a & 150 & Titanium gas diffusion layer \\
b & 150 & Titanium gas diffusion layer after low chemical degradation \\
c & 800 & Titanium gas diffusion layer \\
d & 800 & Titanium gas diffusion layer after low chemical degradation \\
e & 800 & Titanium gas diffusion layer after high chemical degradation \\
f & 15000 & Carbon gas diffusion layer on PTFE substrate \\
g & 15000 & Carbon gas diffusion layer on Nafion substrate \\
h & 15000 & Carbon gas diffusion layer on graphite substrate \\
\end{tabular}
\end{table}

In Figures~\ref{Fig:MatsCorrFuncsA} and \ref{Fig:MatsCorrFuncsB}, eight images from different material samples had been studied.
A label has been asigned for each image, which description is given in Table~\ref{Tab:Materials}.
The eight images (a-h) were obtained from the samples by SEM microscopy, with the same resolution but a particular magnification.

In order to study the effect of the binarization method, two procedures were implemented.
On one hand, binarization had been performed with the commonly used Otsu's method and the corresponding correlation functions had been obtained.
On the other hand, a SVM classification method had been applied to the original image, followed by the calculation of the correlation functions.
In both cases, the correlation functions include the two-point and the line-path functions.
The results are presented in Figures~\ref{Fig:MatsCorrFuncsA} and \ref{Fig:MatsCorrFuncsB}.

Since all the samples come from statistically homogeneous and isotropic materials, their correlation functions show typical trends:
\begin{itemize} 
\item the two-point correlation function takes the value $S_j^{(2)}=\phi$ at $r=0$, followed by a monothonically decreassing path towards the plateau $S_j^{(2)}\phi^2$ at $r\gg0$;
\item the line-path correlation function takes the value $L_j=\phi$ at $r=0$, followed by a monothonically decreassing path towards the plateau $L_j=0$ at $r\gg0$.
\end{itemize}
Therefore, the main features of the above-mentioned correlation functions can be represented only by the surface fraction and the rate at which the curve decays. 

First, one may compare the two-point correlation function $S_j^{(2)}(r)$, for each image (a -- h) and a given phase ($j=0$ or $j=1$).
For instance, the shape of $S_j^{(2)}(r)$ is almost the same for the two binarization methods.
For a given $j$, the difference lies in the value of $\phi_j$, which leads, as a direct consequence, to different leftmost values and plateau levels.
In general, the Otsu's method produces noisy binarized images, mostly in the regions where the solid phase $j=1$ exists, when the materials present a smaller scale roughness.
This can be observed in the treatment of images, mainly for intermediate and large magnifications (c -- h).
A noisy binarized image shows a higher $\phi_1$ for the void phase and a smaller $\phi_0$ for the solid phase than the expected values of the surface fraction: fakes increase and decrease of the void and solid surface fractions, respectively, are generated.
In contrast, the SVM binarization method captures the existence of solid aggregates, despite the roughness of their surfaces, thus yielding more accurate values of $\phi_j$.
For most samples (a, b, d -- h), the difference in $\phi_j$ between the two methods is greater than 0.10.
Sample c shows smooth solid surfaces, which shrinks the difference in $\phi_j$ between the two methods below 0.10.
The opposite extreme case is observed for sample g, where the roughness of the material provokes the misleading disappearance of a significant section of the solid phase, when Otsu's method is applied.
This creates a difference greater than 0.30, between the values of $\phi_j$, obtained with the two methods, for the corresponding phases.  
Another important advantage of the SVM method can be discerned from sample b, where the poor contrast of the original image and, maybe, an inclination of the sample occur.
The Otsu method is sensitive to the above-mentioned sample preparation or experimental faults, whereas the SVM method rectifies and detects precisely the solid phase aggregates.

For the line-path correlation function $L_j(r)$, there are other important issues besides the difference in the leftmost value $L_j(0)=\phi_j$, which is intrinsic to the phase detection.
Otsu's method induces a fast decay of $L_j(r)$ for the solid phase $j=1$, which is due to the noisy binarized images that a simple threshold produces when applied on a wrinkled solid surface.
In contrast and once more, the SVM binarization method reveals the existence of solid aggregates despite the roughness of their surfaces, and, as a consequence, $L_j(r)$ for the solid phase $j=1$ presents a slower decay, indicating the connectedness of the $j=1$ phase through straight lines of solid material.
Additionally, the $L_j(r)$ for the void phase $j=0$ obtained with the SVM method presents a curve with lower values than its Otsu equivalent, and also a less step descent along the entire curve.

\clearpage

\begin{sidewaysfigure}[h!]
\ContinuedFloat*
\centering
\includegraphics[width=1.0\textwidth]{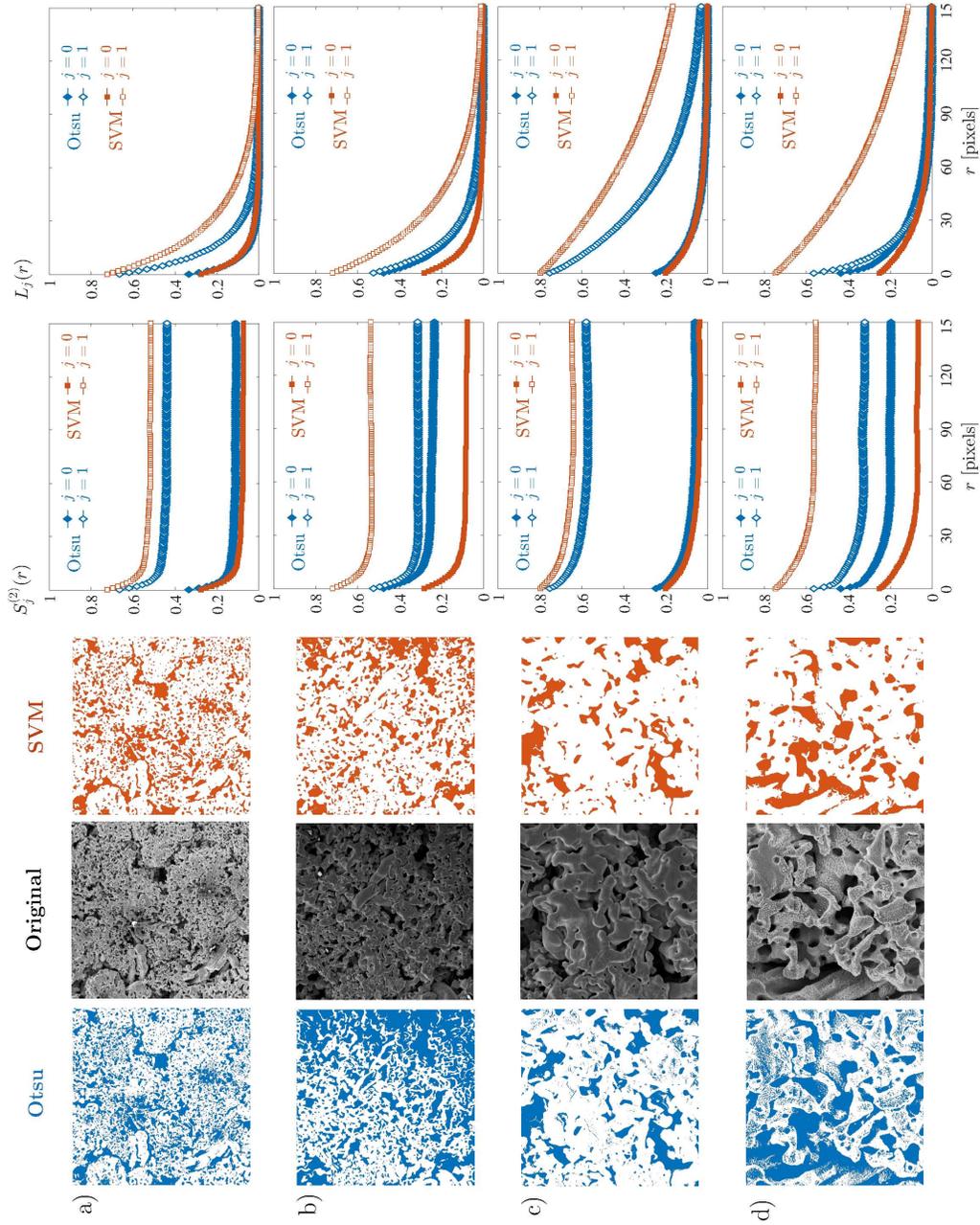}
\caption{Binarization and correlation functions of samples a) -- d), described in Table~\ref{Tab:Materials}. For each type of sample, the columns correspond to (from left to right): Otsu binarization, original image, SVM binarization, two-point correlation function and line-path correlation function. Both correlation functions are shown as functions of the distance $r$, being computed for the two phases ($j=0$ for void and $j=1$ for solid) exhibited by each sample, and with the two binarization methods Otsu and SVM.}
\label{Fig:MatsCorrFuncsA}
\end{sidewaysfigure}

\begin{sidewaysfigure}[h!]
\ContinuedFloat
\centering
\includegraphics[width=1.0\textwidth]{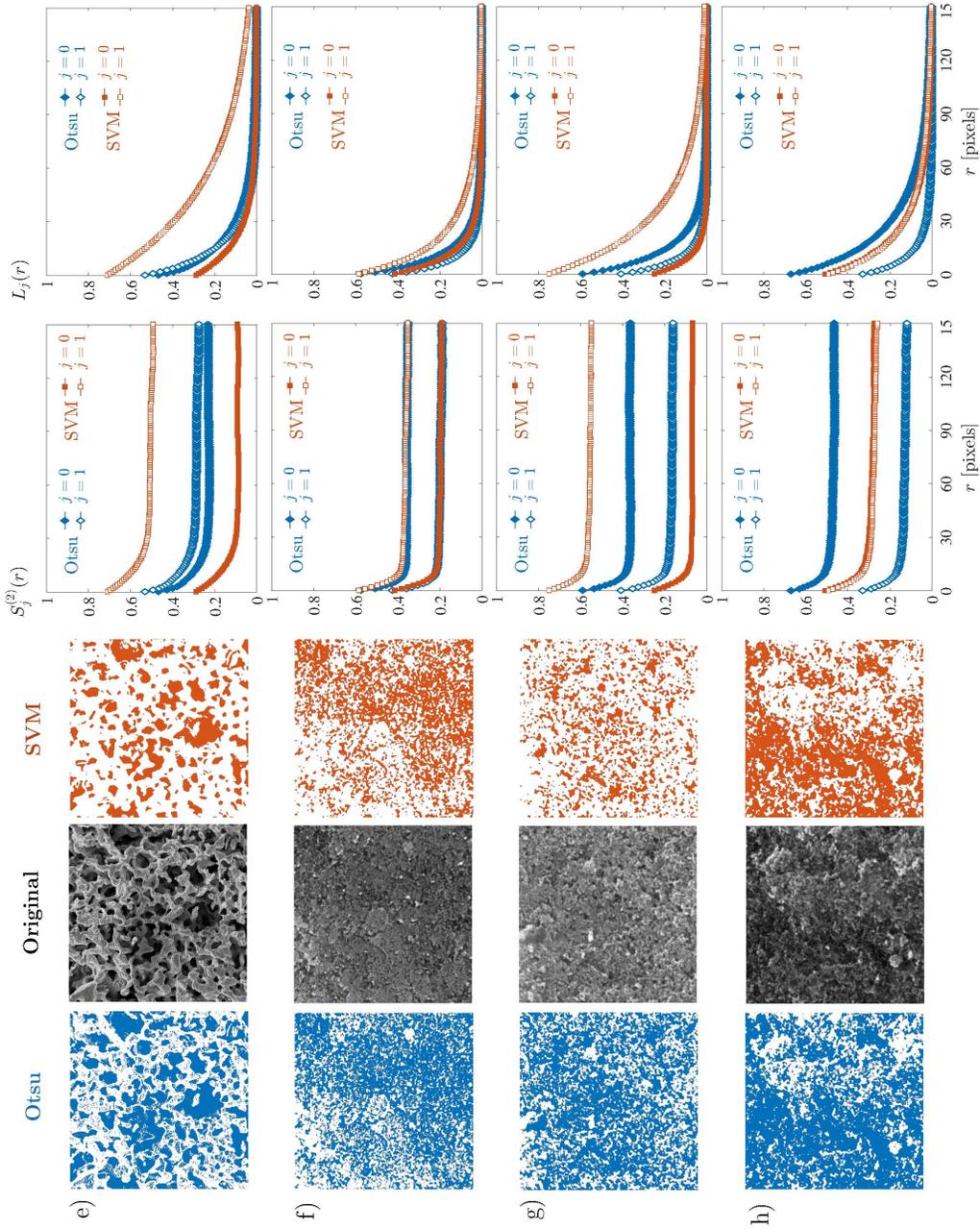}
\caption{Binarization and correlation functions of samples e) -- h), described in Table~\ref{Tab:Materials}. For each type of sample, the columns correspond to (from left to right): Otsu binarization, original image, SVM binarization, two-point correlation function and line-path correlation function. Both correlation functions are shown as functions of the distance $r$, being computed for the two phases ($j=0$ for void and $j=1$ for solid) exhibited by each sample, and with the two binarization methods Otsu and SVM.}
\label{Fig:MatsCorrFuncsB}
\end{sidewaysfigure}

\clearpage

\section{Conclusion}
\label{conclusion}
The pixel's classification of images obtained from random heterogeneous materials is a relevant step to compute their physical properties, like Effective Transport Coefficients (ETC), during a characterization process as stochastic reconstruction. 
The SVM classification method allows to generate a binarized image representing an authentic cut of the material, despite the fact that images are obtained from the surface of the sample.
Additionally, the SVM method allows to include image information further than grayscale pixel values; in this proposal, the image gradient (direction and magnitude) has been used besides the gray intensity.
Even though, high-order microstructure statistical information has not been considered,
the effect of the binarization method is clear from the comparison of the low-order correlation functions for the same material phase.
The SVM methods shows a better agreement with what an observer may discern as void and solid phases from an image of a two-phases material sample. For instance, in the case study, the accuracy of the pixel's classification is 77.6\% for the SVM method and 40.9\% for Otsu's method.

\section*{Acknowledgment}

The authors would like to thank SEP- CONACYT under the grant CB-2013/221988.


\bibliographystyle{elsarticle-num}

\end{document}